\documentclass[us letters]{article}
\usepackage{amsmath,amssymb,graphicx,amssymb,delarray,subfigure,verbatim, bbding,boondox-calo}
\usepackage{diagbox, makecell}
\usepackage{bbding}
\usepackage{multirow}
\usepackage{boondox-calo}

\title{A deep complex multi-frame filtering network for stereophonic acoustic echo cancellation}
\name{Linjuan Cheng$^{1,2,3}$, Chengshi Zheng$^{1,3}$, Andong Li$^{1,3}$, Yuquan Wu$^{2,3}$, Renhua Peng$^{1,3}$, Xiaodong Li$^{1,3}$}
\address{
	$^1$Key Laboratory of Noise and Vibration Research, Institute of Acoustics, Chinese Academy of Sciences, Beijing, China\\
	$^2$Science \& Technology on Integrated Infomation System Laboratory, Institute of Software Chinese Academy of Sciences, Beijing, China\\
	$^3$University of Chinese Academy of Sciences, Beijing, China}
\email{ \{chenglinjuan, cszheng, liandong, pengrenhua, lxd\}@mail.ioa.ac.cn, yuquan@iscas.ac.cn}

\begin{document}
	
	\maketitle
	\begin{abstract}
		In hands-free communication system, the coupling between loudspeaker and microphone generates echo signal, which can severely influence the quality of communication. Meanwhile, various types of noise in communication environments further reduce speech quality and intelligibility. It is difficult to extract the near-end signal from the microphone signal within one step, especially in low signal-to-noise ratio scenarios. In this paper, we propose a deep complex network approach to address this issue. Specially, we decompose the stereophonic acoustic echo cancellation into two stages, including linear stereophonic acoustic echo cancellation module and residual echo suppression module, where both modules are based on deep learning architectures. A multi-frame filtering strategy is introduced to benefit the estimation of linear echo by capturing more inter-frame information. Moreover, we decouple the complex spectral mapping into magnitude estimation and complex spectrum refinement. Experimental results demonstrate that our proposed approach achieves stage-of-the-art performance over previous advanced algorithms under various conditions.
	\end{abstract}
	\section{Introduction}
	In hands-free audio and video communication system, the acoustic coupling between loudspeaker and microphone generates echo signal, which severely reduces the quality of communication. Stereophonic systems are becoming more and more popular because they can provide the listener with spatial information over the single-channel system \cite{benesty2001advances}. Traditional stereophonic acoustic echo cancellation (SAEC) algorithms usually use adaptive filters to identify the echo paths and acoustic echo can then be canceled~\cite{amand1996identifying}. These algorithms are affected by the correlation between stereo far-end signals, resulting in a well known non-unique problem \cite{sondhi1993acoustic, sondhi1995stereophonic}. Thus, the echo cancellation performance degrades rapidly when the echo paths between the stereophonic loudspeakers and the microphone change or the far-end speaker moves from one position to another position. 
	
	To solve the non-unique problem, a series of decorrelation methods have been proposed to reduce the correlation between far-end signals as a preprocessor before implementing SAEC algorithm \cite{benesty1999frequency}. Benesty et al. \cite{1997A} proposed to introduce a small nonlinearity into each channel. Romoli et al. \cite{2010A} utilized the missing fundamental phenomenon to decorrelate the stereophonic channels by suppressing the fundamental frequency component of one far-end signal frame by frame. To further improve the performance, a hybrid decorrelation method has been proposed in \cite{2016Stereophonic}, where an improved sinusoidal phase modulation was applied in the high-frequency band and a pitch-driven sinusoidal injection scheme with a simplified binaural masking model was adopted in the low-frequency band. Although these decorrelation methods can mitigate the nonuniqueness to a certain extent, almost all of them degrade the audio quality and stereophonic spatial perception to some degree. What is more, these SAEC algorithms can usually only remove some linear echo components, and thus some post-processing methods are often necessary to combine with these SAEC algorithms \cite{2001On,Eberhard2000Hands}. A stereophonic acoustic echo suppression (SAES) approach without preprocessing was proposed in \cite{2012Stereophonic}, which employed the Wiener filter in the short-time Fourier transform (STFT) to suppress the echo signal directly instead of identifying echo path directly. An improved SAES algorithm has been proposed in \cite{2014Stereophonic} by incorporating the spectral and temporal correlations in the STFT domain. These SAES algorithms usually perform well in stationary noise scenarios while recude their performance greatly in non-stationary noise scenarios. And their performance is also limited in the presence of double talk and the rapid change of echo path. 
	
	Recently, with the rapid development of deep neural networks (DNNs), DNN-based monophonic acoustic echo cancellation (MAEC) algorithms have become a hot research topic. They can be roughly classified into two categories. One is to combine DNN with traditional adaptive filtering for echo cancellation and the DNN serves as a residual echo suppressor \cite{lee2015dnn,carbajal2018multiple}. The other is to directly suppress the acoustic echo in an end-to-end manner using DNN \cite{zhang2018deep, fazel2019deep}. Deep learning-based MAEC methods are robust to double-talk situation and can better suppress acoustic echo and noise even in non-stationary noise environments. The performance of deep learing-based MAEC methods in real-world environment has been verified in the AEC-Challenge \cite{peng2021icassp, peng2021acoustic}, which has released numerous real recording clips for researchers to test their proposed algorithms in practical acoustic scenarios. 
	
	Nonetheless, deep learing has not received adequate attention in SAEC research yet. A convolutional recurrent network (CRN)-based SAES algorithm has been first proposed in \cite{cheng2021deep}, which used a DNN to directly suppress stereophonic acoustic echo from microphone signal without decorrelation. And a CRN-based complex SAES with a two-stage approach was proposed in \cite{xuebao2021}. The echo signal is estimated in the first stage and the near-end speech is estimated in the second stage by using the estimated echo signal and the microphone signal as the inputs for network. Meanwhile, the input features and training targets used in the network are the complex spectra of signals, which can also recover the phase of the near-end speech. The two-stage complex CRN method performs much better than a single CRN model in both single-talk and double-talk situations and the deep learning-based SAES methods do not have the nonunique problem, thus it is unnecessary to decorrelate the far-end signals. 
	
	However, we found that the two-stage complex CRN algorithm performs worse in lower signal-to-noise ratio (SNR) scenarios. Using a single-stage network to estimate complex spectrum has proved inadequate in relatively difficult tasks due to its limited mapping capability \cite{xuebao2021}. Recently, the adavantage of multi-stage training has been revealed in many tasks, e.g., speech separation and enhancement. To improve the performance of SAES algorithms in relative low SNRs, this paper proposes a multi-stage approach to decompose the echo reduction and complex spectrum mapping problems. Firstly, we divide the echo reduction into the linear echo cancellation and residual echo suppression. A DNN-based multi-frame filtering structure is used to preprocess the linear echo and a stacked network is used to suppress residual echo and noise. Secondly, estimation of the complex spectrum of the near-end speech is decomposed into the amplitude and phase estimation networks. In other words, when estimating the near-end speech, there are two separate networks. One is to estimate the spectral magnitude of the near-end speech, the other is to refine the phase spectrum.
	
	The remainder of this paper is organized as follows. Section 2 introduces the signal model and formulates the problem. The overall framework of the proposed algorithm is described in Section 3. The experimental setup is presented in Section 4. In Section 5, the results and analysis are given to validate the performance of the proposed framework. Finally, some conclusions are drawn in Section 6.
	\begin{figure}[t]
		\centering
		\includegraphics[width=0.95\linewidth]{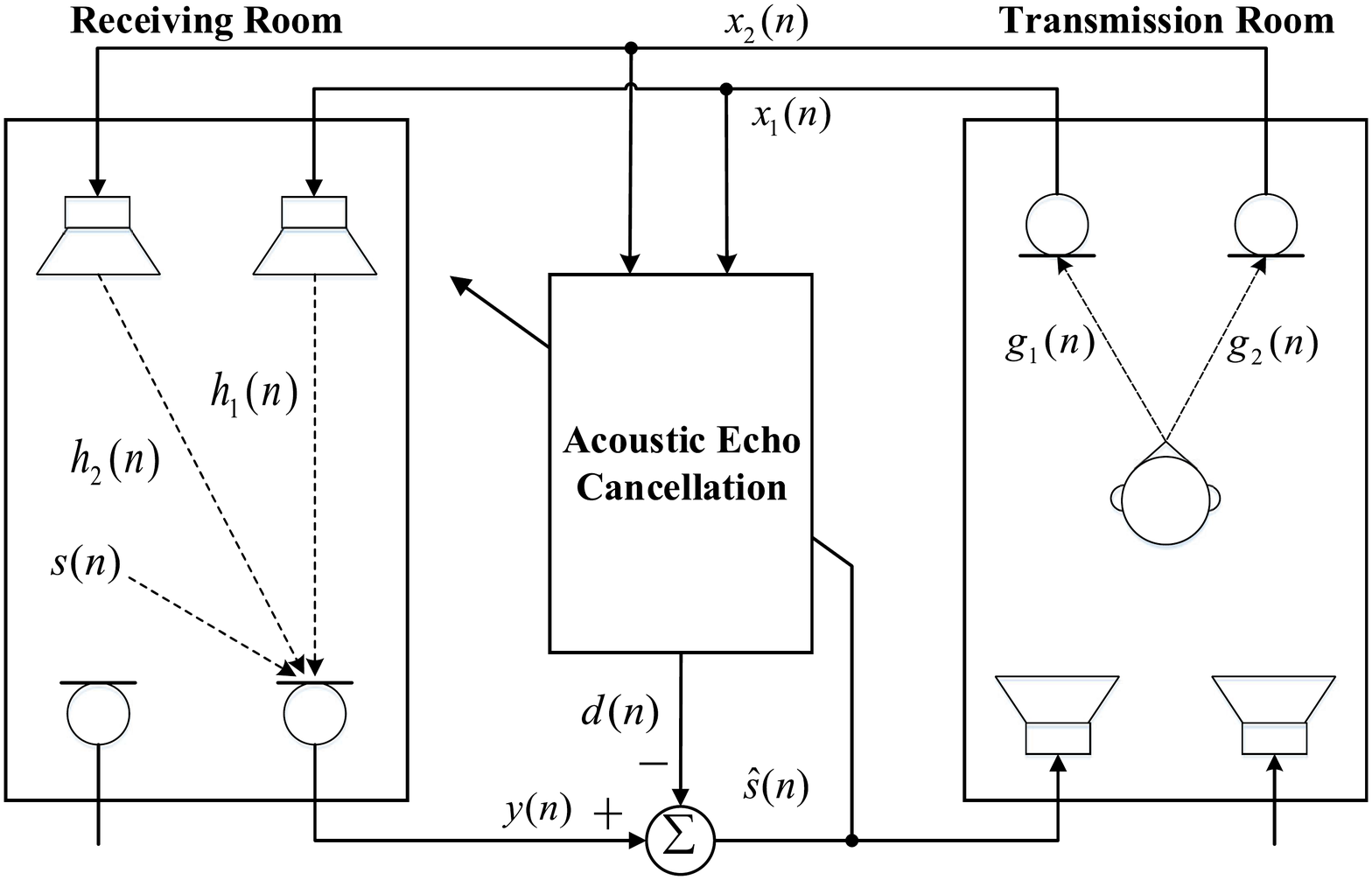}
		\caption{Diagram of a stereophonic acoustic echo system.}
		\label{fig:SAEC Diagram}
		\vspace{-0.8cm}
	\end{figure}
	\vspace{-0.4cm}
	\section{Signal Model}
	\vspace{-0.2cm}
	The diagram of a typical SAEC system is illustrated in Fig.~{\ref{fig:SAEC Diagram}}. The receiving room is on the left while the transmission room is on the right. Without loss of generality, we choose only one microphone in the receiving room to formulate the SAEC problem and the same framework can be extended to the other microphone. In the tranmission room, the two far-end signals $x_{1}(n)$ and $x_{2}(n)$ are generated by a common source $r(n)$ via room impluse reponses (RIRs) $g_{1}(n)$ and $g_{2}(n)$. In the receiving room, the transmitted far-end signals $x_{1}(n)$ and $x_{2}(n)$ are played by two loudspeakers and coupled with one of the microphones via the acoustic paths denoted by RIRs $h_{11}(n)$ and $h_{12}(n)$. Accordingly, the microphone signal can be modeled as:
	\begin{gather}
	\begin{aligned}
	\label{eqn1}
	y_{1}(n)&=x_{1}(n)\ast h_{11}(n)+x_{2}(n)\ast h_{12}(n)+s_{1}(n)+v_{1}(n)\\
	&=d_{1}(n)+s_{1}(n)+v_{1}(n),
	\end{aligned}
	\end{gather}
	where ${\ast}$ denotes linear convolution, $n$ denotes the discrete time index, $d_{1}(n)$ is the echo signal, $s_{1}(n)$ is the near-end speech signal, and $v_{1}(n)$ represents the additive environmental noise. The goal of SAEC is to estimate the near-end signal ${s}_{1}(n)$ from $y_{1}(n)$ with the two far-end signals $x_{1}(n)$ and $x_{2}(n)$. 
	\vspace{-0.4cm}
	\section{Algorithm description}
	\vspace{-0.2cm}
	\begin{figure*}[t]
		\centering
		\includegraphics[width=0.6\linewidth]{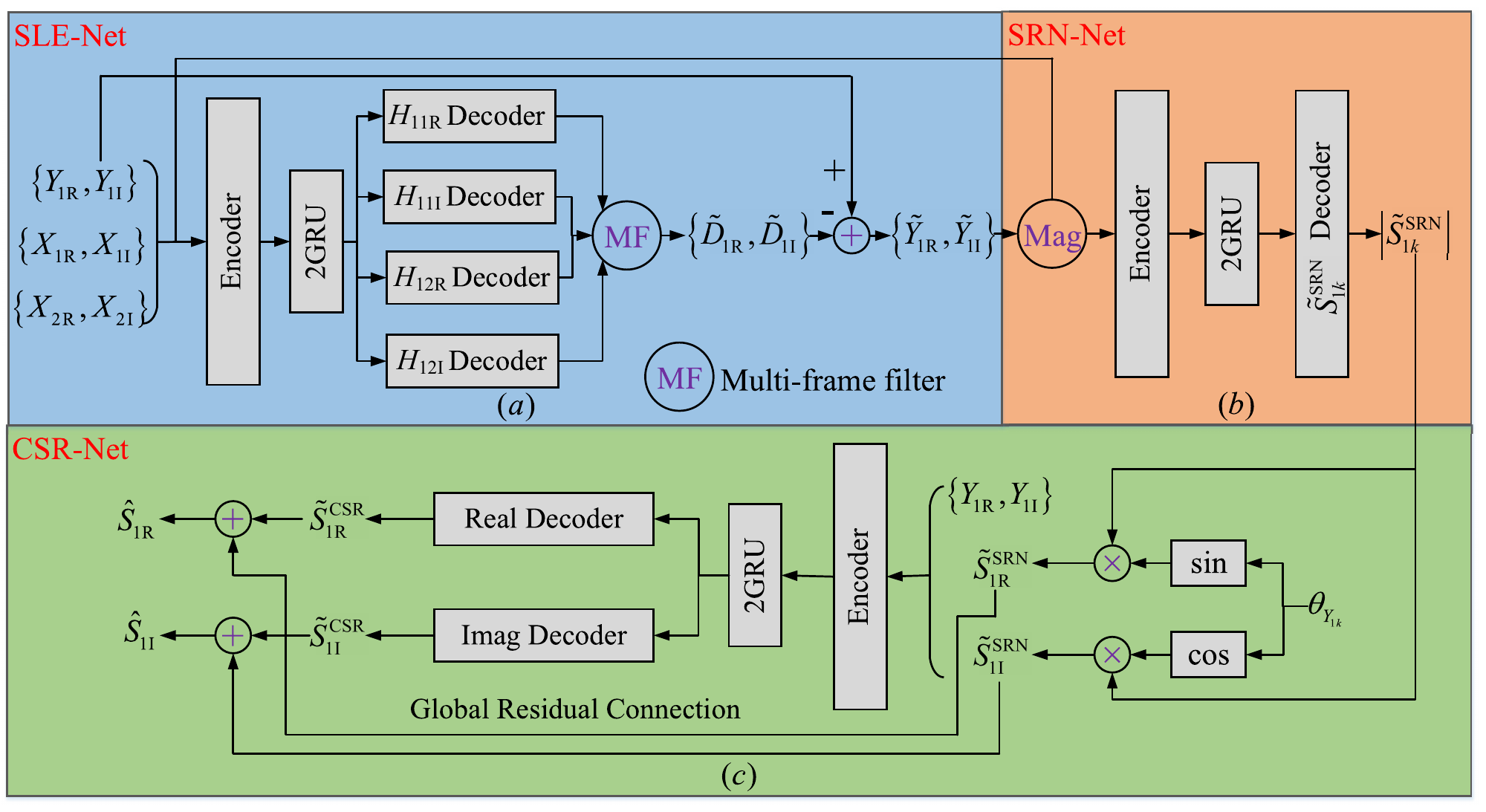}
		\caption{Overall diagram of the proposed framework.}
		\vspace{-0.6cm}
		\label{fig:Proposed algorithm scheme}
	\end{figure*}
	\subsection{Network Structure}
	\vspace{-0.2cm}
	The overall framework of the proposed system is depicted in Fig.~{\ref{fig:Proposed algorithm scheme}}. We denote 
	$\{ Y_{1\rm R},Y_{1\rm I} \}$, $\{ X_{1\rm R},X_{1\rm I} \}$, and $\{ X_{2\rm R},X_{2\rm I} \}$
	as the real and imaginary (RI) parts of the complex spectra of $y_{1}(n)$, $x_{1}(n)$ and $x_{2}(n)$, respectively. $\lvert Y_{1k}\rvert$, $\lvert X_{1k}\rvert$, and $\lvert X_{2k}\rvert$ are their corresponding spectral magnitude of $y(n)$, $x_{1}(n)$ and $x_{2}(n)$, respectively. $\tilde{Y}_{1k}$ is the output of SLE-Net containing residual echo and noise. $\{\tilde{S}_{1\rm R}^{\rm{SRN}},\tilde{S}_{1\rm I}^{\rm{SRN}}\}$ are the RI parts of the estimated near-end speech complex spectrum processed by SRN-Net and $\lvert\tilde{S}_{1k}^{\rm{SRN}}\rvert$ is its spectral magnitude. The RI parts of the final estimated near-end speech processed by CSR-Net are notated as $\hat{S}_{1\rm R}$ and $\hat{S}_{1\rm I}$, respectively. Note that the frame index and the frequency index are omitted when no confusion arises.
	
	The proposed architecture consists of three modules, namely suppressing linear echo network (SLE-Net), suppressing residual echo and noise network (SRN-Net), and complex spectrum refining network (CSR-Net). For each module, we adopt the typical CRN structure, which has an encoder, a decoder, and two gated recurrent unit (GRU) layers. The motivation of the proposed algorithm is based on two points. One is that we divide stereophonic echo suppression into two steps, where the SLE-Net aims to estimate the linear echo part while the SRN-Net is designed for suppressing residual echo and background noise. The other is that we decompose the complex spectrum mapping into spectral magnitude estimation and complex spectrum refinement. The output of SRN-Net is the spectral magnitude of the near-end speech. Then a coarse complex spectrum is obtained by coupling the estimated sectral magnitude and the phase of the microphone signal. In CSR-Net, both the coarsely estimated complex spectrum of the near-end speech and the complex spectrum of the microhpne signal are used as inputs to estimate the residual complex spectrum of the near-end speech, which is then added to the coarsely estimated complex spectrum to obtain a refined counterpart. On the one hand, the estimated spectral magnitude reduces the solution space of the complex spectrum optimization network. In addition, the step-by-step approach can improve the performance in complex acoustic scenarios.
	
	The details of SLE-Net are presented in Fig.~{\ref{fig:Proposed algorithm scheme}}(a). The SLE-Net uses the complex spectra of $y(n)$, $x_{1}(n)$, and $x_{2}(n)$ as input features with a dimension of $[6, T, F]$, where $T$ and $F$ denote the size in the time and frequency axes, respectively. The outputs are the complex spectra of estimated RIRs. In Fig. {\ref{fig:Proposed algorithm scheme}}(b), $\{ H_{11\rm R}, H_{11\rm I}\}$ and $\{ H_{12\rm R}, H_{12\rm I}\}$ denote the RI parts of $h_{11}$ and $h_{12}$, respectively. And $\{ \tilde{H}_{11\rm R}, \tilde{H}_{11\rm I}, \tilde{H}_{12\rm R}, \tilde{H}_{12\rm I}\}$ are the corresponding estimated values of $\{ H_{11\rm R}, H_{11\rm I}, H_{12\rm R}, H_{12\rm I}\}$. The estimated RIRs are combined with the two far-end signals in complex spectrum domain to estimate the linear echo components $\{ \tilde{D}_{1\rm R},\tilde{D}_{1\rm I} \}$ as follws:
	\begin{gather}
	\begin{aligned}
	\label{eqn2}
	\tilde{D}_{1\rm R}=f_{\rm MF}(\tilde{H}_{11\rm R}, X_{1\rm R})-f_{\rm MF}(\tilde{H}_{11\rm I}, X_{1\rm I})+\\
	f_{\rm MF}(\tilde{H}_{12\rm R}, X_{2\rm R})-f_{\rm MF}(\tilde{H}_{12\rm I}, X_{2\rm I}),
	\end{aligned}
	\end{gather}
	\begin{gather}
	\begin{aligned}
	\tilde{D}_{1\rm I}=f_{\rm MF}(\tilde{H}_{11\rm R}, X_{1\rm I})+f_{\rm MF}(\tilde{H}_{11\rm I}, X_{1\rm R})+\\
	f_{\rm MF}(\tilde{H}_{12\rm R}, X_{2\rm I})+f_{\rm MF}(\tilde{H}_{12\rm I}, X_{2\rm R}),
	\end{aligned}
	\end{gather}
	where $f_{\rm MF}$ denotes the multi-frame fiter. There is a strong temporal correlation between adjacent frames of echo signal. In generally, the echo path is too long to be covered by only one frame, and thus several adjacent frames of echo signals are often highly correlated. To model this temporal correlation, a multi-frame structure was proposed in [21] to better estimate the linear echo signal \cite{li2021simultaneous}. The sizes of $\{ \tilde{H}_{11\rm R}, \tilde{H}_{11\rm I}, \tilde{H}_{12\rm R}, \tilde{H}_{12\rm I}\}$ are all $[L,T,F]$, where $L$ is the number of channels and it also indicates the filter length. Take $f_{\rm MF}(\tilde{H}_{11\rm R}, X_{1\rm R})$ as an example, it can be given by:
	\begin{gather}
	\begin{aligned}
	f_{\rm MF}(\tilde{H}_{11\rm R}, X_{1\rm R})(l,k)=\sum_{q=0}^{L-1}\tilde{H}_{11\rm R}(q,l,k)\cdot X_{1\rm R}(l-q,k).
	\end{aligned}
	\end{gather}
	
	Then the estimated echo is subtracted from the microphone signal $\{ Y_{1\rm R},Y_{1\rm I} \}$ to obtain $\{ \tilde{Y}_{1\rm R},\tilde{Y}_{1\rm I} \}$ containing residual echo and environmental noise. 
	
	As shown in Fig.~{\ref{fig:Proposed algorithm scheme}}(b), the inputs of SRN-Net include the spectral magnitude of the microphone signal, the two far-end signals, and the output of SLE-Net with a dimension of $[4, T, F]$. The output is the estimated spectral magnitude of the near-end speech, denoted by $\lvert \tilde{S}_{1k}^{\rm SRN} \rvert$.
	
	Fig.~{\ref{fig:Proposed algorithm scheme}}(c) illustrates the detailed stream of CSR-Net. First, $\lvert \tilde{S}_{1k}^{\rm SRN} \rvert$ is combined with the phase of $y_{1}(n)$ to obtain a coarsely estimated complex spectrum of the near-end speech $\{\tilde{S}_{1\rm R}^{\rm SRN},\tilde{S}_{1\rm I}^{\rm SRN} \}$. Second, $\{\tilde{S}_{1\rm R}^{\rm SRN},\tilde{S}_{1\rm I}^{\rm SRN} \}$ are concatenated with $\{Y_{1\rm R},Y_{1\rm I}\}$ as the inputs of CSR-Net. Two decoders in CSR-Net output the RI parts of the residual complex spectrum of the near-end speech, respectively, denoted by $\tilde{S}_{1\rm R}^{\rm CSR}$ and $\tilde{S}_{1\rm I}^{\rm CSR}$. The final estimated near-end speech is obtained by the sum of $\{\tilde{S}_{1\rm R}^{\rm CSR},\tilde{S}_{1\rm I}^{\rm SRN}\}$ and $\{\tilde{S}_{1\rm R}^{\rm SRN},\tilde{S}_{1\rm I}^{\rm SRN} \}$, given by:
	\begin{gather}
	\hat{S}_{1\rm R}=\tilde{S}_{1\rm R}^{\rm CSR}+\tilde{S}_{1\rm R}^{\rm SRN},\\
	\hat{S}_{1\rm I}=\tilde{S}_{1\rm I}^{\rm SRN}+\tilde{S}_{1\rm I}^{\rm SRN}.
	\end{gather}
	
	\renewcommand\arraystretch{0.85}
	\begin{table*}[t]
		\caption{Evaluation results of different methods in terms of PESQ, ESTOI, and ERLE at different SNRs. $\textbf{BOLD}$ denotes the best result in each case.}
		\centering
		\large
		\resizebox{0.92\textwidth}{!}{
			\begin{tabular}{c|c|ccccccccccccccc}
				\toprule
				\hline
				\multirow{2}{*}{Noise}&\multirow{2}{*}{Model}&\multicolumn{5}{c|}{PESQ}&\multicolumn{5}{c|}{ESTOI}&\multicolumn{5}{c}{ERLE}\\
				\cline{3-17}
				& &10\rm{dB} &15\rm{dB} &20\rm{dB} &25\rm{dB} &\multicolumn{1}{c|}{30\rm{dB}}&10\rm{dB} &15\rm{dB} &20\rm{dB} &25\rm{dB} &\multicolumn{1}{c|}{30\rm{dB}}&10\rm{dB} &15\rm{dB} &20\rm{dB} &25\rm{dB} &30\rm{dB}\\
				\hline
				\multirow{5}{*}{Home}&None&2.21&2.35&2.44&2.48&\multicolumn{1}{c|}{2.51}&0.58&0.67&0.74&0.79&\multicolumn{1}{c|}{0.82}&--&--&--&--&--\\
				&NLMS&2.02&2.29&2.51&2.67&\multicolumn{1}{c|}{2.78}&0.52&0.62&0.72&0.81&\multicolumn{1}{c|}{0.87}&2.74&4.61&6.82&9.08&10.88\\
				&Wiener&2.41&2.61&2.75&2.83&\multicolumn{1}{c|}{2.88}&0.62&0.71&0.80&0.85&\multicolumn{1}{c|}{0.88}&2.66&3.89&5.17&6.26&7.05\\
				&CRN&2.90&3.09&3.22&3.29&\multicolumn{1}{c|}{3.33}&0.78&0.84&0.88&0.91&\multicolumn{1}{c|}{0.93}&45.22&43.44&42.21&41.32&40.63\\	
				&TS-CRN&2.95&3.16&3.31&3.38&\multicolumn{1}{c|}{3.43}&0.79&0.85&0.89&0.92&\multicolumn{1}{c|}{0.94}&46.77&45.64&44.47&43.28&42.53\\
				&TS-CCRN&2.99&3.21&3.37&3.45&\multicolumn{1}{c|}{3.51}&0.79&0.85&0.90&0.92&\multicolumn{1}{c|}{0.94}&42.77&41.91&40.93&40.08&39.56\\
				&SLE-SRN&3.02&3.27&3.45&3.56&\multicolumn{1}{c|}{3.62}&0.79&0.86&0.90&0.93&\multicolumn{1}{c|}{0.95}&45.31&44.31&43.47&42.80&42.39\\
				&Proposed&\textbf{3.11}&\textbf{3.35}&\textbf{3.52}&\textbf{3.63}&\multicolumn{1}{c|}{\textbf{3.69}}&\textbf{0.84}&\textbf{0.89}&\textbf{0.92}&\textbf{0.95}&\multicolumn{1}{c|}{\textbf{0.96}}&\textbf{52.89}&\textbf{51.13}&\textbf{49.68}&\textbf{48.55}&\textbf{47.67}\\
				\hline
				\multirow{5}{*}{Babble}&None&1.94&2.04&2.08&2.11&\multicolumn{1}{c|}{2.12}&0.54&0.61&0.65&0.68&\multicolumn{1}{c|}{0.69}&--&--&--&--&--\\
				&NLMS&1.73&1.97&2.09&2.16&\multicolumn{1}{c|}{2.19}&0.50&0.63&0.75&0.83&\multicolumn{1}{c|}{0.89}&4.52&7.32&10.47&13.71&16.23\\
				&Wiener&2.14&2.29&1.36&2.40&\multicolumn{1}{c|}{2.41}&0.63&0.72&0.79&0.83&\multicolumn{1}{c|}{0.85}&5.08&7.23&9.15&10.66&11.59\\
				&CRN&2.64&2.82&2.90&2.95&\multicolumn{1}{c|}{2.97}&0.76&0.82&0.85&0.87&\multicolumn{1}{c|}{0.88}&43.46&43.06&42.68&42.18&41.76\\
				&TS-CRN&2.68&2.91&3.01&3.07&\multicolumn{1}{c|}{3.09}&0.77&0.83&0.87&0.89&\multicolumn{1}{c|}{0.90}&44.91&44.92&44.33&43.61&43.16\\
				&TS-CCRN&2.75&2.98&3.09&3.16&\multicolumn{1}{c|}{3.18}&0.77&0.84&0.87&0.89&\multicolumn{1}{c|}{0.90}&41.06&40.71&40.26&39.90&39.56\\
				&SLE-SRN&2.75&2.97&3.09&3.15&\multicolumn{1}{c|}{3.18}&0.79&0.86&0.90&0.92&\multicolumn{1}{c|}{0.93}&43.95&43.52&43.00&42.46&42.07\\
				&Proposed&\textbf{2.85}&\textbf{3.07}&\textbf{3.19}&\textbf{3.25}&\multicolumn{1}{c|}{\textbf{3.29}}&\textbf{0.84}&\textbf{0.89}&\textbf{0.92}&\textbf{0.94}&\multicolumn{1}{c|}{\textbf{0.95}}&\textbf{52.98}&\textbf{51.77}&\textbf{50.76}&\textbf{49.85}&\textbf{49.21}\\
				\hline
				\bottomrule
		\end{tabular}}
		\label{tbl:pesq-comparison}
		\vspace*{-0.45cm}
	\end{table*}
	\vspace{-0.4cm}
	\subsection{Loss function}
	The output of SLE-Net is a finite-length estimated RIR, which is regraded as a hidden mapping in this paper. The training of this network includes two stages. In the first stage, the parameters of SRN-Net and SLE-Net are updated by using the same cost function, which is the mean square error (MSE) of the spectral magnitude of the near-end speech, given by:
	\begin{equation}
	\begin{aligned}
	\label{eq:MSE} 
	\mathcal{L}_{\rm {SLE-SRN}}& =\left( \lvert \tilde{S}_{1k}^{\rm SRN} \rvert - \lvert S_{1k}\rvert\right)^2\\
	& = \Big ( M_{1}^{\rm SRN}\lvert \tilde{S}_{1k}^{\rm SRN} \rvert- \lvert S_{1k}\rvert \Big )^2,
	\end{aligned}
	\end{equation}
	where $M_{1}^{\rm SRN}$ is the output of the last layer in the decoder of SRN-Net and $\lvert S_{1k}\rvert$ is the spectral magnitude of the near-end speech $s_{1}(n)$. Here we use a signal approximation (SA) method to estimate the spectral magnitude of near-end speech.
	
	After training SLE-Net and SRN-Net, in the second stage, the complex spectrum of the near-end speech is used to train the CSR-Net, while the parameters of SLE-Net and SRN-Net are refined to get better optimization. The loss function in the second stage can be given by:
	\begin{gather}
	\mathcal{L}_{\rm {CSR}} =\tilde{\mathcal{L}}_{\rm {CSR}}+0.1\mathcal{L}_{\rm {SLE-SRN}},\\
	\tilde{\mathcal{L}}_{\rm {CSR}}= 0.5\left( \hat{S}_{1R}-S_{1R} \right)^{2}+0.5\left( \hat{S}_{1I}-S_{1I} \right)^{2}.
	\end{gather}
	
	\section{Experimental Results}
	\subsection{Experimental setup}
	The TIMIT corpus \cite{zue1990speech} is chosen to evaluate the performance of the proposed approach. 120 pairs of speakers are randomly selected as the far-end and near-end speakers. Each speaker has ten sentences. For each far-end speaker, three utterances are randomly selected and concatenated to form a far-end signal. An utterance from a near-end speaker is randomly chosen and extended to the same length as that of the far-end signal by zeros-padding both in front and in rear. The RIRs are generated using the image method~\cite{allen1979image}. Three sizes of near-end and far-end rooms are selected, which are $[4,3,3]$ m, $[6,4,3]$ m, and $[8,7,3]$ m, respectively. The reverberation time is randomly selected from $\left\{0.3 s,  0.6 s, 0.9 s\right\}$. The distance between loudspeakers is $2.0$ m and the distance between microphones is $0.4$ m. The distance between each speaker position and the center of microphones is set to be $[0.3,0.7,1.1]$ m. The near-end speech is mixed with the generated echo signals at a signal-to-echo ratio (SER) randomly chosen from $[0, 5, 10, 15]$ dB. $881$ non-stationary noises~\cite{snyder2015musan} are used as background noise to be mixted with the near-end speech and generated echo signal at SNR = $[10, 15, 20, 25, 30]$ dB. The total amount of the training dataset is about $100$ hours. We randomly choose $80\%$ of these mixtures for training and the remaining $20\%$ are chosen for validation.
	
	All of these signals are sampled at $16$ kHz. Take the microphone signal $y(n)$ as an example, we use a 320-point (20 ms) hamming window to segment $y_{1}(n)$ with 50\% overlap in consecutive frames. 320-point STFT is applied, resulting in a 161-dimensional spectral feature in each frame. Each encoder includes five layers with the numbers of channels being $[8, 16, 32, 64, 128]$. Accordingly, the numbers of channels of five layers in each decoder are $[64, 32, 16, 8, 1]$. The kernel size and stride of enconders and decoders are $(1,3)$ and $(1,2)$ in the time and frequency axes except that the kernel size is set to $(3,3)$ in SLE-Net to utilize the inter-frame information. The networks are optimized by Adam algorithm~\cite{kingma2015adam}. The learning rate is set to $3\times10^{-4}$ and the mini-batch size is 16 at the utterance level. 
	
	Two typical traditional algorithms including a NLMS-based SAEC \cite{yang2016stereophonic} and a Wiener filter-based SAES \cite{yang2012stereophonic} are chosen as the baselines. Meanwhile, the proposed framework is also compared with three advanced deep learning-based SAES methods, which are a single CRN model (CRN) \cite{cheng2021deep}, a two-stage CRN model with spectral magnitude mapping (TS-CRN) and complex spectrum mapping (TS-CCRN)~\cite{xuebao2021}. In the proposed approach, the filter length $L$ is set to $10$. To evaluate the performance of decomposing amplitude and phase strategy, the SLE-Net and SRN-Net are adopted to directly estimate the complex spectrum of the near-end speech and it is set as a baseline, namely SLE-SRN. In the single-talk case, ERLE is used as the measurement metric to evaluate the echo attenuation, while PESQ \cite{rix2001perceptual} and ESTOI \cite{jensen2016algorithm} are used to evaluate speech quality and intelligibility in the double-talk case.
	\subsection{Results and Analysis}
	\begin{figure}[t]
		\centering
		\includegraphics[width=1.0\linewidth]{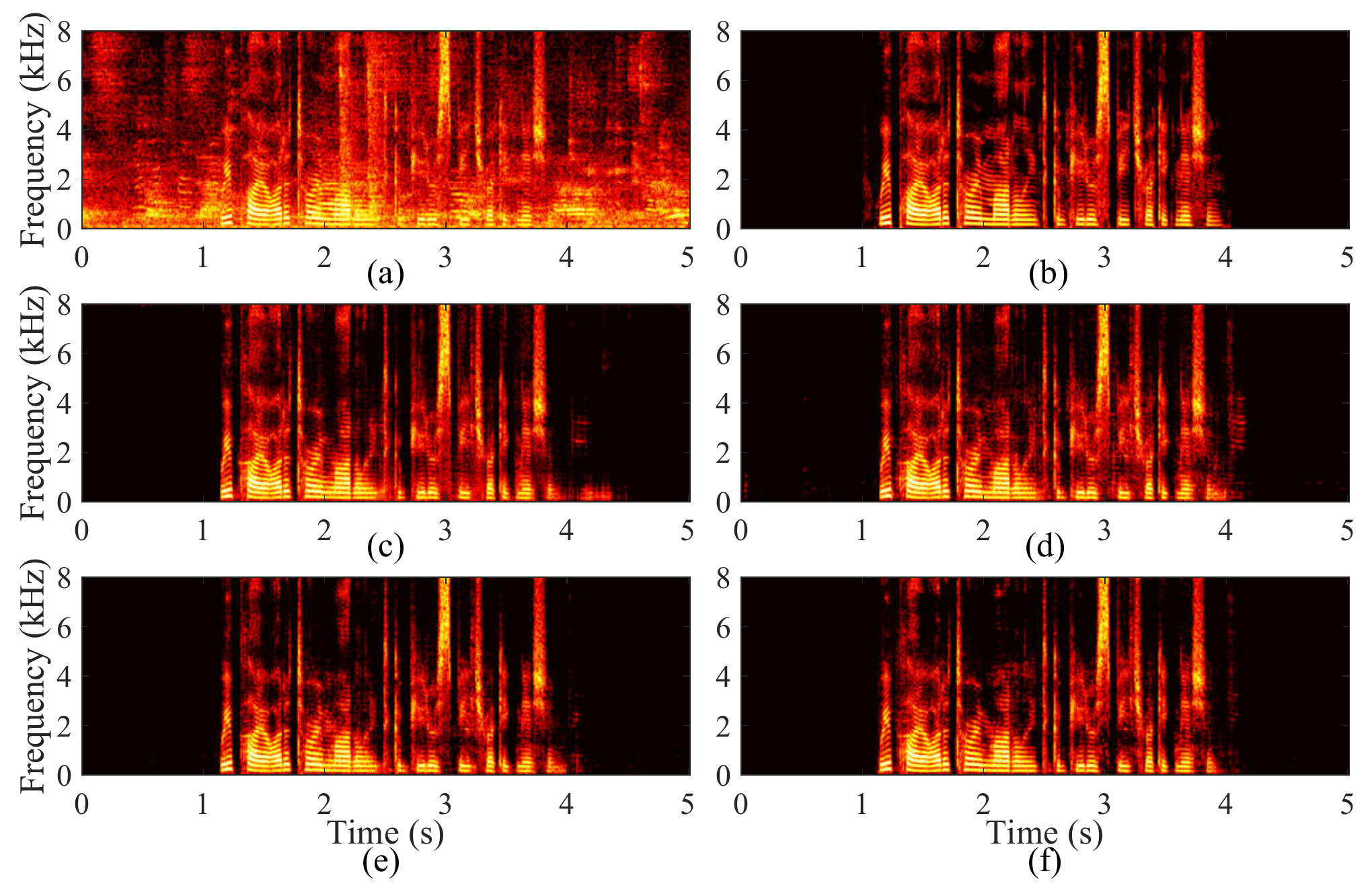}
		\caption{Spectrogram of enhanced near-end speech processed by different algorithms. (a) microphone signal, PESQ = 2.08; (b) near-end signal; (c) processed by TS-CRN algorithm, PESQ = 2.78; (d) processed by TS-CCRN algorithm, PESQ = 2.85; (e) processed by SLE-SRN algorithm, PESQ = 2.89; (f) processed by Proposed algorithm, PESQ = 2.96.}
		\label{fig:spectrogram}
		\vspace{-0.4cm}
	\end{figure}
	
	The test results of different methods in terms of PESQ, ESTOI, and ERLE are summarized in Table 1. The background noises include Home noise from QUT-NOISE dataset \cite{dean2010qut} and Babble noise from Noise-92 dataset \cite{varga1993assessment}. Traditional NLMS and Wiener methods perfom worse in non-stationary noise scenarios in both single-talk and double-talk situations. The performance of the single CRN model is much better than traditional methods and its ERLE results can achieve more than $40$ dB under different SNR conditions, indicating the advantage of deep learning method. The proposed algorithm can achieve the best performance in all conditions, especially in low SNR cases. The ERLE scores of TS-CCRN method are lower than TS-CRN, indicating that the mapping complex spectrum directly does not benefit the echo suppression. Introducing multi-frame structure can improve the ERLE score by about $3$ dB, and the proposed algorithm can further improve the ERLE by approximately $7$ dB. The ESTOI scores of TS-CRN and TS-CCRN are almost the same. However, the ESTOI score difference between the proposed algorithm and TS-CCRN is about $0.03$ in Home noise and $0.05$ in Babble noise.
	
	The spectrograms of the microhpne signal, the near-end signal, and the enhanced signals are plotted in Fig. {\ref{fig:spectrogram}}. It is shown that there are still some residual echo components in the enhanced signal processed by TS-CRN. Compared with TS-CCRN and SLE-SRN, the proposed method can better suppress the residual echo while reducing speech distortion, especially in low-frequency bands. 
	\section{Conclusions}
	This paper proposes a deep complex multi-frame filtering network for stereophonic acoustic echo cancellation. In the proposed framework, linear echo is first estimated using SLE-Net with a multi-frame filtering structure while the residual echo and noise is processed by SRN-Net. In order to better extract the phase information of the near-end speech, the complex spectrum of the near-end speech is estimated by decoupling its amplitude and phase recovery. The experimental results show that the proposed framework improves the speech quality and intelligibility in double-talk situations and the echo and noise can be better suppressed in single-talk situations. 
	\bibliographystyle{IEEEtran}
	%\bibliography{mybib}

	% \begin{thebibliography}{9}
	% \bibitem[1]{Davis80-COP}
	%   S.\ B.\ Davis and P.\ Mermelstein,
	%   ``Comparison of parametric representation for monosyllabic word recognition in continuously spoken sentences,''
	%   \textit{IEEE Transactions on Acoustics, Speech and Signal Processing}, vol.~28, no.~4, pp.~357--366, 1980.
	% \bibitem[2]{Rabiner89-ATO}
	%   L.\ R.\ Rabiner,
	%   ``A tutorial on hidden Markov models and selected applications in speech recognition,''
	%   \textit{Proceedings of the IEEE}, vol.~77, no.~2, pp.~257-286, 1989.
	% \bibitem[3]{Hastie09-TEO}
	%   T.\ Hastie, R.\ Tibshirani, and J.\ Friedman,
	%   \textit{The Elements of Statistical Learning -- Data Mining, Inference, and Prediction}.
	%   New York: Springer, 2009.
	% \bibitem[4]{YourName17-XXX}
	%   F.\ Lastname1, F.\ Lastname2, and F.\ Lastname3,
	%   ``Title of your INTERSPEECH 2021 publication,''
	%   in \textit{Interspeech 2021 -- 20\textsuperscript{th} Annual Conference of the International Speech Communication Association, September 15-19, Graz, Austria, Proceedings, Proceedings}, 2020, pp.~100--104.
	% \end{thebibliography}
	
\end{document}